\documentclass[%
 aip,
cp,  
 amsmath,amssymb,
 reprint,%
]{revtex4-2}

\usepackage{graphicx}
\usepackage{dcolumn}
\usepackage{bm}

\usepackage[utf8]{inputenc}
\usepackage[T1]{fontenc}
\usepackage{mathptmx} 

\begin{document}

\title{Meson-Nucleon Scattering Amplitudes from Lattice QCD}

\author{John Bulava} 
 \email{bulava@cp3.sdu.dk}
\affiliation{
  CP3-Origins, University of Southern Denmark, Campusvej 55, 5230 
Odense M, Denmark}

\date{\today} 

\begin{abstract}
Lattice QCD calculations of resonant meson-meson scattering 
amplitudes have improved significantly due to algorithmic and computational 
advances. However, progress in meson-nucleon scattering has been slower due to
difficulties in computing the necessary correlation functions, the exponential
signal-to-noise problem, and the finite-volume treatment of scattering with 
fermions. Nonetheless, first benchmark calculations have now been performed. The status of lattice QCD calculations of 
meson-nucleon scattering amplitudes is reviewed together with comments on future prospects. 
\end{abstract}

\maketitle

Meson-nucleon scattering amplitudes are required in a variety of active research areas of nuclear and particle physics. Near threshold 
they provide valuable determinations of the low-energy constants (LEC's) of chiral perturbation theory involving baryons,
which also govern the long-range (pion-exchange) component of the inter-nucleon forces.
The first principles calculation of meson-nucleon amplitudes from lattice QCD both compliments experiment and
 probes the quark mass dependence, 
enabling studies of the convergence of the asymptotic perturbative series. Low-energy pion-nucleon scattering is also related 
to the pion-nucleon sigma term $\sigma_{\pi N}$~\cite{Cheng:1970mx,Brown:1971pn}, which is an input to the phenomenology of 
dark matter direct-detection experiments. Some tension currently exists between lattice determinations of 
$\sigma_{\pi N}$ and phenomenological 
extractions of $\sigma_{\pi N}$ from $\pi N$ scattering data~\cite{RuizdeElvira:2017stg}, so that precise lattice calculations 
of the $\pi N$ scattering lengths are needed.
At somewhat larger center-of-mass energy, electroweak form-factors of the $N\rightarrow \Delta(1232) \rightarrow N\pi$ 
transition are needed for neutrino-nucleus scattering experiments~\cite{Alvarez-Ruso:2017oui}, and the nature of the 
low-lying $N(1440)$ and $\Lambda(1405)$ resonances remain unsettled.     

Although lattice QCD is well established as a reliable non-perturbative approach to calculating the properties of single 
hadrons,
lattice QCD studies of hadron scattering amplitudes have proven more difficult. Increased development of the formalism 
for relating finite-volume lattice observables to infinite-volume amplitudes, the rapid growth of computer power, 
and novel efficient numerical algorithms have combined to advance the state-of-the-art. This review highlights the substantial 
recent progress achieved in lattice QCD calculations of hadron scattering amplitudes, in particular on the nascent subfield 
of meson-nucleon systems. A related review of lattice QCD calculations of meson-baryon scattering amplitudes is given in 
Ref.~\cite{Morningstar:2019jjz}. 

A detailed introduction to lattice QCD is beyond the scope of this work~\cite{Gattringer:2010zz,DeGrand:2006zz}, but 
several main points are worthy of mention. Lattice QCD 
employs the path 
integral formulation of QFT regulated on a finite   
space-time lattice of $(L/a)^3\times(T/a)$ points, where $a$ denotes the lattice spacing and $L$ ($T$) the spatial (temporal) 
extent. After analytically integrating out the quark fields, Markov chain Monte Carlo methods sample the remaining 
path integral over the gluon field (denoted $U$) to produce stochastic estimates for Euclidean 
$n$-point correlation functions. Employing the time-momentum representation for a correlation function between hadron
interpolating fields ${\cal O}_i$ and ${\cal O}_j$ results in the spectral decomposition 
\begin{align}\label{e:ct} 
C_{ij}^{\rm 2pt}(\boldsymbol{p},\tau) = \langle {\cal O}_i(\boldsymbol{p},\tau) \, {\cal O}^{\dagger}_j(0) \rangle_U  = 
\sum_{n} \langle 0 | \hat{\cal O}_i(\boldsymbol{p}) | n \rangle \, \langle n | \hat{\cal O}^{\dagger}_j(0) | 0 \rangle \, 
{\rm e}^{-E_n \tau} \, + \, {\rm O}({\rm e}^{-MT})
\end{align} 
where $\langle \dots \rangle_U$ denotes the Monte Carlo average over suitably distributed gluon fields and the sum runs over 
finite-volume Hamiltonian eigenstates $\hat{H}|n\rangle = E_n | n\rangle$. The form of the exponentially suppressed 
finite-$T$ effects depends on the temporal boundary conditions and the interpolating operators. Evidently the large-time limit
 of these Euclidean $n$-point functions is dominated by the lowest contributing eigenstates and analyses of 
$C^{\rm 2pt}_{ij}(\boldsymbol{p},\tau)$ treating the first few terms in the sum enable a determination of the low-lying 
finite-volume 
energies and matrix elements~\cite{Michael:1985ne,Luscher:1990ck,Blossier:2009kd}. Such determinations are hampered by 
the exponential signal-to-noise degradation typically present as $\tau\rightarrow \infty$ so that in practice 
finite-volume levels are isolated successfully only if the employed set of operators $\{{\cal O}_i\}$ has 
significant overlap onto them.  

As is evident from Eq.~\ref{e:ct}, lattice QCD simulations are performed in Euclidean time $\tau = it$. The usual determination 
of scattering amplitudes 
in real time relies on an asymptotic formalism, such as the well-known LSZ~\cite{LSZ} and Haag-Ruelle~\cite{Haag,Ruelle} approaches, 
where the asymptotic time limits $t\rightarrow \pm \infty$ of correlation functions isolate the desired in and out states. 
Although an analogous asymptotic 
formalism has been proposed for Euclidean time in Ref.~\cite{Barata:1990rn}, Maiani and Testa~\cite{Maiani:1990ca} proved 
that the naive large-separation limit of Euclidean correlators does not (in general) yield on-shell scattering amplitudes. 
Fortunately, a work-around was developed by L\"{u}scher~\cite{Luscher:1990ux} in which the finite-volume is used as a tool 
to probe hadron interactions. In this approach the signal consists of the 
deviation of finite-volume two-hadron levels from their non-interacting values. Although originally formulated for 
total momentum $\boldsymbol{P}=0$ 
elastic scattering between two spinless identical particles, generalizations have been developed for
moving frames~\cite{Rummukainen:1995vs,Kim:2005gf,Fu:2011xz}, non-identical particles with 
spin~\cite{Gockeler:2012yj,Briceno:2014oea,Morningstar:2017spu}, asymmetric volumes~\cite{Feng:2004ua}, 
coupled two-hadron scattering channels~\cite{He:2005ey,Lage:2009zv,Bernard:2010fp,Briceno:2012yi,Hansen:2012tf}, 
and amplitudes containing external currents~\cite{Lellouch:2000pv,Lin:2001ek,Detmold:2004qn,Meyer:2011um,Bernard:2012bi,
Briceno:2012yi,Hansen:2012tf,Detmold:2014fpa,Agadjanov:2014kha,Feng:2014gba,
Briceno:2014uqa,Briceno:2015csa,Briceno:2015tza,Baroni:2018iau}.
Although this approach, which is reviewed in Ref.~\cite{Briceno:2017max}, is applicable to two-to-two scattering only below three (or more) 
hadron thresholds, extensions treating three-hadron amplitudes are under 
development~\cite{Briceno:2012rv,
Polejaeva:2012ut,
Hansen:2014eka,
Meissner:2014dea,
Hansen:2015zga,
Briceno:2017tce,Mai:2017bge,Hammer:2017uqm,Hammer:2017kms,Doring:2018xxx,Mai:2018djl,
Briceno:2018aml,Romero-Lopez:2018rcb,Hansen:2019nir,Blanton:2019igq}.  

Using the notation of Ref.~\cite{Morningstar:2017spu}, the relation between a finite-volume two-hadron energy $E^{\Lambda}_L(\boldsymbol{P})$ and 
infinite-volume two-to-two scattering amplitudes is given by 
\begin{align}\label{e:det}
 \det[K^{-1}(E_{\rm cm}) - B^{(\boldsymbol{P})}(L,E_{\rm cm})] + {\rm O}({\rm e}^{-ML}) = 0 , \qquad E_{\rm cm} =  
\sqrt{(E^{\Lambda}_L(\boldsymbol{P}))^2 + \boldsymbol{P}^2} 
\end{align}
where the exponentially suppressed corrections are ignored in practical applications and $\Lambda$ denotes an irreducible 
representation (irrep) of the relevant finite-volume symmetry group. This `quantization condition' yields information about
the infinite-volume $K$-matrix in the form of a determinant over all possible total angular momenta ($J$), total spin 
combinations ($S$), and two-hadron scattering channels ($a$). Encoding the reduced symmetry of the cubic volume in lattice QCD 
simulations,
the (known) $B$-matrix mixes the infinite number of possible $J$ while the $K$-matrix is diagonal in $J$ but dense in $S$ and 
$a$. The practical application of Eq.~\ref{e:det} requires a block diagonalization in the basis of finite-volume irreps, and a 
truncation to the limited set of contributing partial waves below some maximum orbital angular momentum $\ell_{\rm max}$. 
Only in the single-channel, single-partial wave approximation is there a one-to-one correspondence between 
$E^{\Lambda}_L(\boldsymbol{P})$ and $K^{-1}(E_{\rm cm})$. In other cases a global fit is performed to all energies 
resulting in a model-dependent parametrization of $K^{-1}(E_{\rm cm})$. 

We turn now to the lattice QCD calculation of finite-volume two-hadron energies $\{E^{\Lambda}_L(\boldsymbol{P})\}$ from
the two-point correlation functions in Eq.~\ref{e:ct}. 
Correlation functions
between two-hadron interpolating operators in which each hadron has definite spatial momentum are most effective in
isolating two-hadron states. However, the measurement of such correlation functions on an ensemble of 
gauge field configurations has long been a computational challenge. Since the Grassmann-valued quark fields are integrated 
out analytically in the lattice QCD path integral, Wick's theorem is employed to express hadron correlation functions in terms 
of quark propagators. On a finite discrete Euclidean lattice, the quark propagators between space-time points $x$ and $y$ is 
given as the inverse of the large, sparse,ill-conditioned Dirac operator $M^{-1}(x,y)$. This matrix is so large that its inverse 
computed only by solving linear systems $M\phi = \eta$ for some `sink' $\phi(x)$ given a `source' $\eta(y)$. 
Projection onto definite spatial momentum however requires knowledge of the entire matrix inverse, so-called 
all-to-all quark propagators. The calculation of all required elements of this matrix inverse on each gauge configuration 
by solving one linear system for each space-time point is prohibitively expensive, 
so alternative algorithms are required. One approach that has been particularly successful is the 
Laplacian-Heaviside method (LapH)~\cite{Peardon:2009gh,Morningstar:2011ka} in which the all-to-all quark propagator is 
projected onto the 
subspace spanned by the $N_{\rm ev}$ lowest modes of the gauge covariant Laplace operator, reducing its dimension 
significantly and enabling the computation of the smeared all-to-all propagator. This method has the added benefits of affecting
a form of quark smearing, whereby the overlap onto the high-lying states in Eq.~\ref{e:ct} is suppressed, and reducing
the problem of 
correlation function construction to the contraction of individual hadron tensors, for which significant optimizations 
can be applied~\cite{Horz:2019rrn}. Although the LapH approach still requires 
a large number of inversions on each gauge configuration, advances in algorithms~\cite{Luscher:2007se,Babich:2010qb} 
for solving linear systems involving the Dirac matrix $M$ have improved the computational cost significantly.  

Complimentary developments in the finite-volume formalism and in the computation of two-hadron energies in lattice QCD 
have driven recent progress in calculations of two-hadron scattering amplitudes. Although much of this progress 
has been in amplitudes with two pseudoscalar mesons, first meson-nucleon calculations have been performed. It is 
these 
calculations that are highlighted here, with work on threshold $N\pi$ amplitudes summarized separately from first 
calculations of the low-lying $\Delta(1232)$ resonance. Finally, a new approach for determining scattering amplitudes 
for spectral functions \emph{without} employing the finite volume is reviewed, before concluding with a summary and 
future prospects.   

\section{Threshold Scattering amplitudes}\label{s:thresh}

The natural first application of the methods described above is to near-threshold meson-nucleon amplitudes, where the 
effective range expansion of $\ell=0$ partial waves 
\begin{align}\label{e:eff}
p_{\rm cm} \cot \delta_0(p_{\rm cm}) = -\frac{1}{a_0} + \frac{1}{2}r_0p_{\rm cm}^2 + P_0r_0^3 p_{\rm cm}^4 + 
{\cal O}(p_{\rm cm}^6) 
\end{align}
defines the scattering length $a_0$, the effective range $r_0$, and the shape parameter $P_0$. These are of course 
parametrization-independent constants that characterize the low-energy interaction and thus of phenomenological significance.
On the finite-volume formalism side, truncating Eq.~\ref{e:det} to $s$-wave contributions only yields a one-to-one 
correspondence between lattice QCD energies and scattering amplitudes. Although not required in practical applications,
the threshold expansion of Eq.~\ref{e:det} is instructive and yields
\begin{align}\label{e:de}
  \Delta E \equiv E^{G_{1g}}_{mN}(L) - M_m - M_N = \frac{-2\pi a_0}{\mu_{mN} L^3} \left[ 1 + c_1 \frac{a_0}{L} + 
  c_2\left(\frac{a_0}{L}\right) \right] + {\rm O}\left(\frac{1}{L}\right)^6    
\end{align}
where $E^{G_{1g}}_{mN}(L)$ is the finite-volume energy of the lowest lying meson-nucleon energy in the $G_{1g}$ irrep,
$\mu_{mN}$ the reduced mass of the meson-nucleon system,  
and $c_1$ and $c_2$ are known constants~\cite{Luscher:1986pf}. Eq.~\ref{e:de} illustrates that finite-volume energy shifts indeed 
constitute the signal for scattering amplitudes, and that sensitivity to higher-order terms in Eq.~\ref{e:eff} occurs only
at ${\rm O}\left(\frac{1}{L}\right)^6$. 

\begin{figure}
\includegraphics[width=0.54\textwidth]{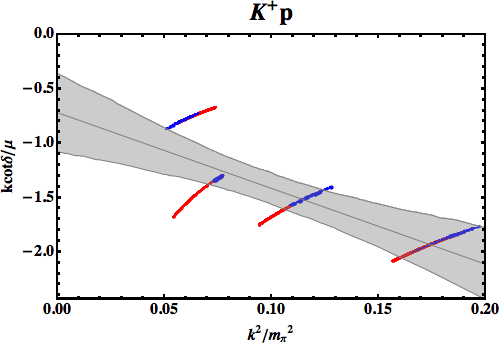}
\includegraphics[width=0.45\textwidth]{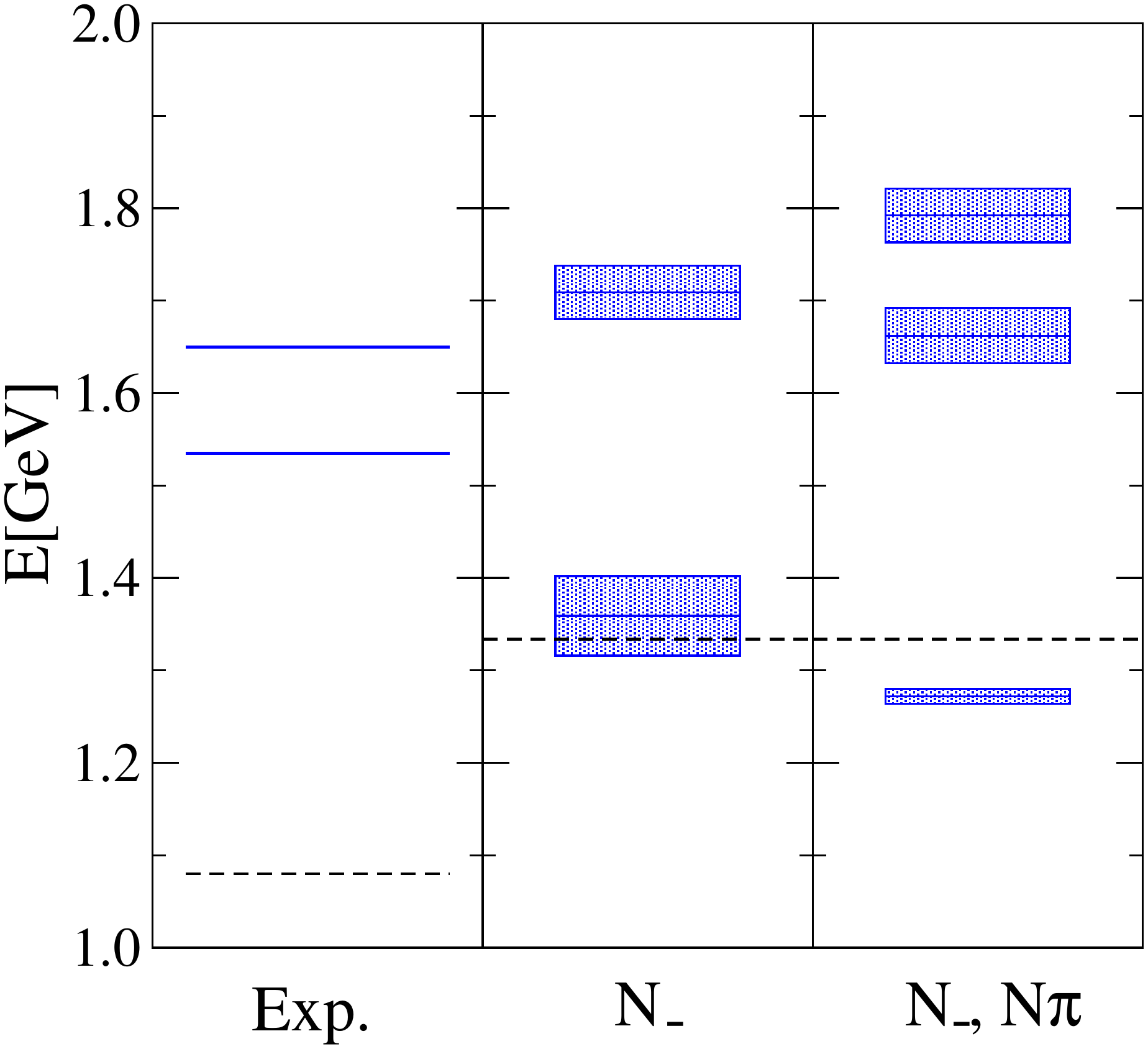}
\caption{\label{f:thresh}\emph{Left}: the low-energy $K^+p$ scattering amplitude from Ref.~\cite{Detmold:2015qwf} on a single 
ensemble with $N_{\rm f}=2+1$ dynamical Wilson fermions at pion mass $m_{\pi}=390{\rm MeV}$. The red and 
blue points are results on individual bootstrap samples for two different analysis strategies and the gray band is a fit to 
Eq.~\ref{e:eff} including the first two terms in the effective range expansion. \emph{Right}: low-lying finite volume 
spectrum in the $I=1/2$ $G_{1u}$ irrep from Ref.~\cite{Lang:2012db} on a single $N_{\rm f}=2$ ensemble with $m_{\pi}=266{\rm MeV}$. 
The first column are the experimentally known $N^*(1535)$ and $N^*(1650)$ 
resonances and the second column is the lattice QCD finite volume spectrum determined using only single-hadron 
operators. The third column is the spectrum determined accurately using both single-hadron and $N\pi$ interpolating 
operators, illustrating the need for a complete basis in practical calculations. }
\end{figure}

Several low-energy $s$-wave meson-nucleon scattering amplitudes have been calculated using this approach in 
Refs.~\cite{Torok:2009dg,Detmold:2015qwf}. The quantum numbers of the overall system are chosen to 
ensure the absence of `same-time' diagrams which contain quark propagators that 
 start and end at the same time. Since only a single ground state is
 determined, projecting onto 
definite momentum everywhere is not strictly required, although it does enhances the overlap onto the desired level.   
Refs.~\cite{Torok:2009dg,Detmold:2015qwf} exploit these simplifications and do not employ the LapH approach for the required 
quark propagators. Ref.~\cite{Detmold:2015qwf}
employs ensembles of $N_{\rm f}=2+1$ dynamical fermions with a single lattice spacing and 
pion masses $m_{\pi} = 390, \, 250{\rm MeV}$. The results for low-energy $K^+p$ scattering  are shown in 
Fig.~\ref{f:thresh}.  

Ref.~\cite{Lang:2012db} computes the lowest three $I=1/2$ $N\pi$ finite-volume energy levels in the $G_{1u}$ irrep for which 
 the leading partial wave approximation yields the near-threshold $s$-wave amplitude. Since this system has non-maximal 
 isospin, same-time diagrams are required and the LapH method is used to efficiently determine the (smeared) all-to-all
  quark propagators. Definite three-momentum projection is performed for all hadrons and 
  appropriate interpolators for each of the three lowest-lying states are included. Results for the three 
  finite-volume energy levels for a single ensemble of $N_{\rm f} = 2$ dynamical fermions with $m_{\pi}=266{\rm MeV}$ are 
  shown in Fig.~\ref{f:thresh}. In this analysis, single-hadron interpolating operators for the lowest-lying resonances in 
  this channel are also included. The importance of appropriate interpolating operators is illustrated by comparison with 
  the lattice spectra obtained with and without the $N\pi$ operator. Due to the exponentially degrading signal-to-noise ratio, there is a 
   finite time range over which the signal can be tracked and the lowest-lying level is determined 
    correctly only if the $N\pi$ is included. Although only the first level is relevant for the near-threshold scattering 
    amplitude, the next two are in qualitative agreement with the experimentally determined $N^*(1535)$ and 
    $N^*(1650)$ resonances.

\section{Resonant amplitudes}

As is evident in Fig.~\ref{f:thresh}, both single-hadron and two-hadron interpolating operators are required to determine the 
low-lying spectrum in the presence of resonances. Below three or more hadron thresholds, this spectrum can be interpreted 
according to Eq.~\ref{e:det} to obtain information about meson-nucleon scattering amplitudes above threshold. 
The benchmark example of such an analysis is the $I=3/2$ $p$-wave elastic $N\pi$ scattering amplitude, which 
contains the narrow $\Delta(1232)$ resonance. This analysis must include single-hadron $\Delta(1232)$ interpolating 
operators as well as $N\pi$ operators with appropriate overall quantum numbers. 

\begin{figure}
\includegraphics[width=0.54\textwidth]{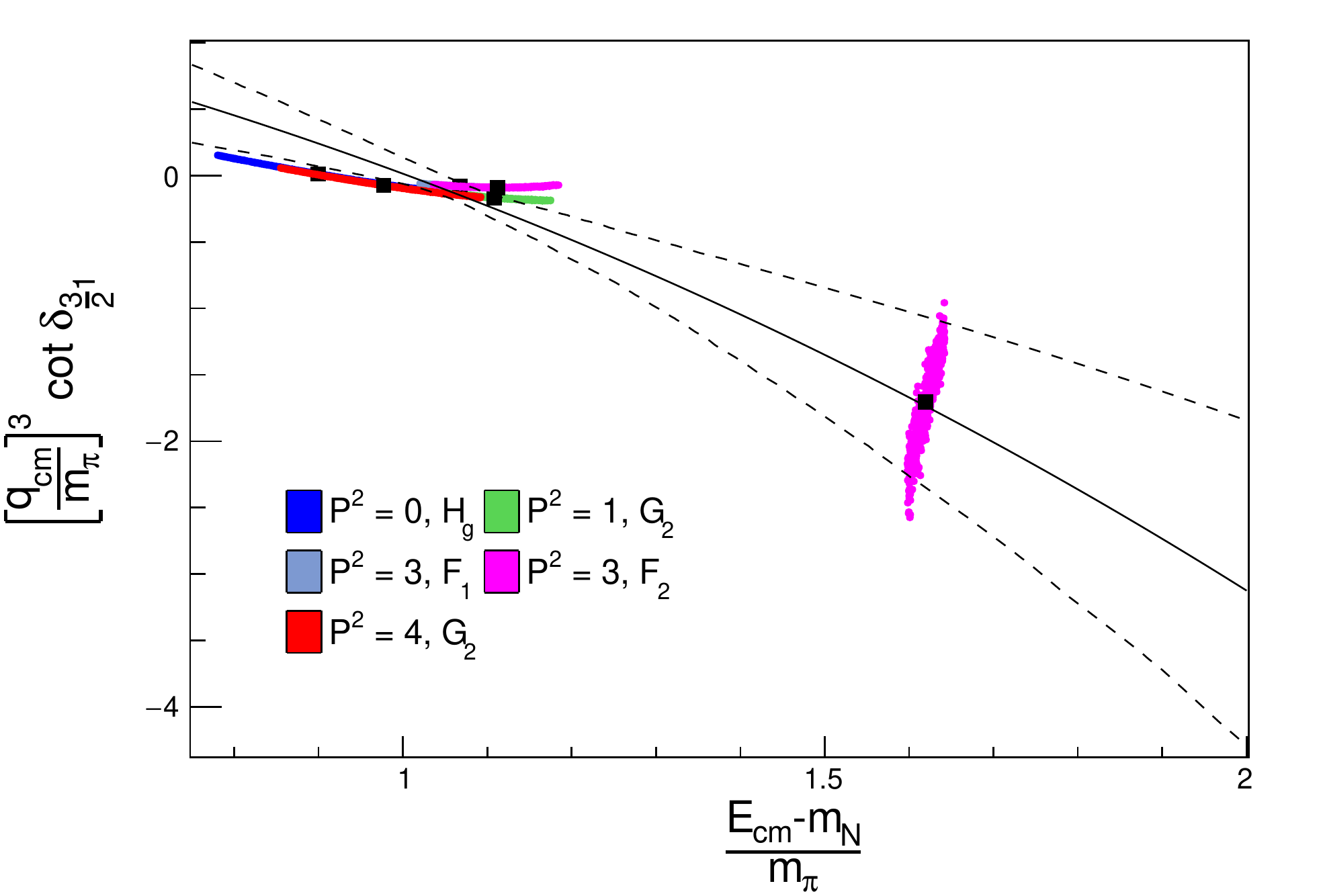}
\includegraphics[width=0.45\textwidth]{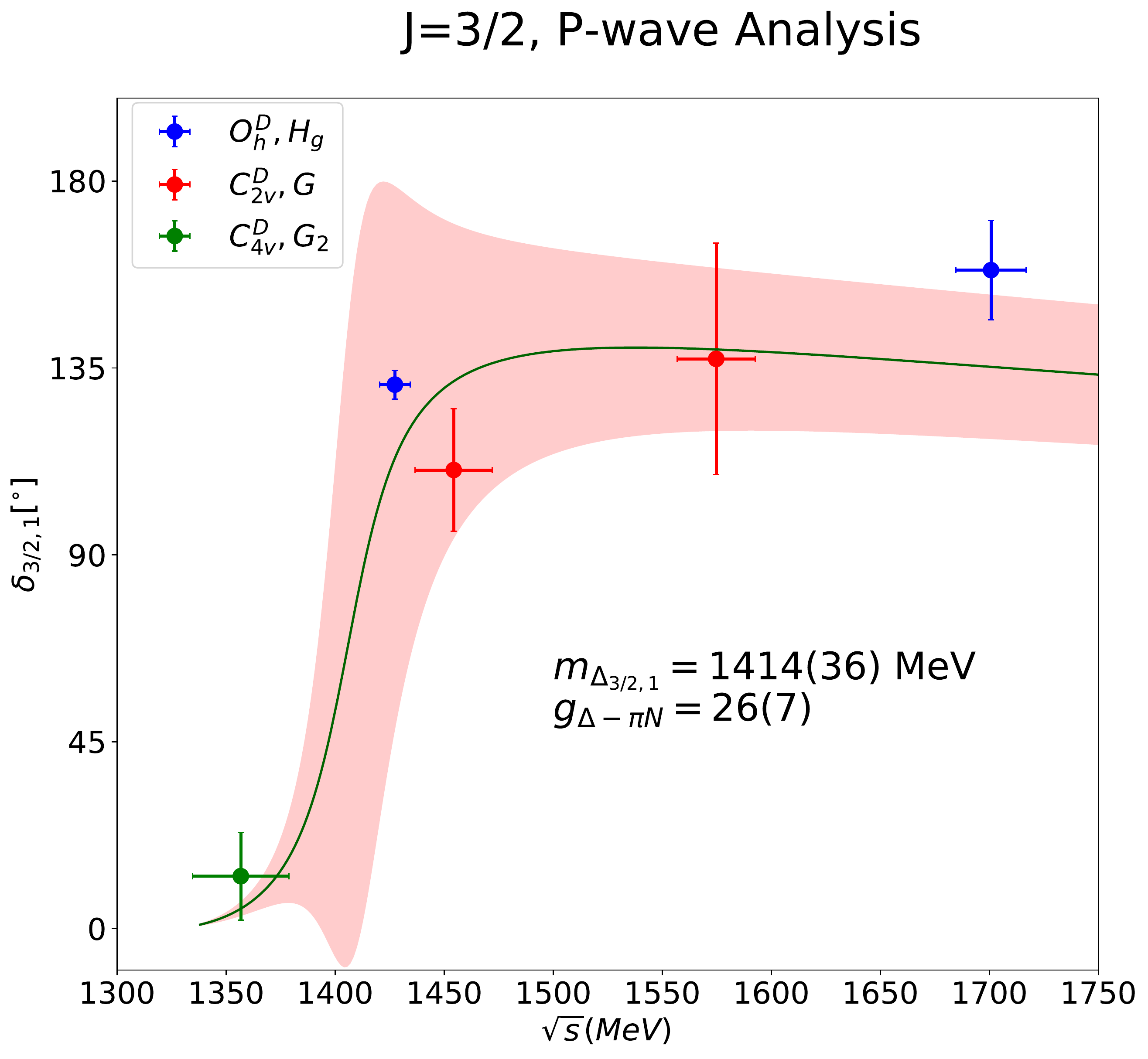}
\caption{\label{f:delta} \emph{Left}: the elastic $I=3/2$ $p$-wave $N\pi$ scattering amplitude from Ref.~\cite{Andersen:2017una}
on a single ensemble of $N_{\rm f}=2+1$ Wilson fermions with $m_{\pi}=280{\rm MeV}$. Due to the unphysical values of the 
quark masses the $\Delta(1232)$ resonance is located near the $N\pi$ threshold. Individual bootstrap samples are shown for each 
point to illustrate correlations and the lines indicate a Breit-Wigner fit. Ref.~\cite{Andersen:2017una} also 
justified the single-partial wave approximation by expanding the determinant condition in Eq.~\ref{e:det} to include all contributing 
$d$-waves, finding their contribution negligible. \emph{Right}: Preliminary $I=3/2$ $p$-wave $N\pi$ scattering phase shift from 
Ref.~\cite{Paul:2018yev} on a single ensemble of $N_{\rm f} = 2+1$ dynamical fermions at $m_{\pi} = 250{\rm MeV}$. The shaded 
region indicates the error band from a Breit-Wigner fit.}
\end{figure}

The first published results on the $\Delta(1232)$ resonance appeared in Ref.~\cite{Andersen:2017una}, although a report on 
preliminary earlier work is found in Ref.~\cite{Mohler:2012nh}. Ref.~\cite{Andersen:2017una} employs a single ensemble of 
$N_{\rm f}=2+1$ dynamical quarks at $m_{\pi} = 280{\rm MeV}$ which lies on a quark mass trajectory that fixes 
$2m_{\rm u,d} + m_{\rm s} = {\rm const.}$ to the physical value. Because the degenerate light quark masses are larger than 
their physical values, the $\Delta(1232)$ is approximately stable and located near $E_{\rm thresh} = m_{N} + m_{\pi}$. This hampers a determination 
of the energy dependence of the amplitude, as is evident in Fig.~\ref{f:delta}. Nonetheless, a Breit-Wigner fit to $p^3 \cot \delta$ 
yields $m_{\Delta} = 1344(20){\rm MeV}$ and $g^{\rm BW}_{\Delta N\pi} = 19.0(7.4)$ which are consistent with phenomenological determinations
from experimental data. Although Ref.~\cite{Andersen:2017una} employs the single partial wave approximation, the influence of 
$d$-waves is checked explicitly by enlarging the determinant in Eq.~\ref{e:det} using the formulae and computer programs published
in Ref.~\cite{Morningstar:2017spu}.    

Preliminary results on the $\Delta(1232)$ resonance are also found in Ref.~\cite{Paul:2018yev} on a single ensemble of 
$N_{\rm f} = 2+1$ dynamical fermions at $m_{\pi}=250{\rm MeV}$. There the leading partial wave approximation is applied 
 and all necessary Wick contractions are evaluated using a combination of sequential, 
stochastic and point-to-all propagators. Preliminary results from Ref.~\cite{Paul:2018yev} are also shown in Fig.~\ref{f:delta}, 
which result in a value of the coupling $g_{\Delta N\pi}$ compatible with Ref.~\cite{Andersen:2017una} and the preliminary determination in 
Ref.~\cite{Verduci:2014btc}. 

In both Refs.~\cite{Andersen:2017una} and~\cite{Paul:2018yev} energies are calculated in a number of kinematic frames, but with 
finite-volume irreps judiciously chosen so that $\ell=1$ is the lowest contributing infinite-volume partial wave. In a 
number of other irreps, both $s$- and $p$-waves contribute, enabling a larger number of constraints on the $K$-matrix due to 
the increased number of  
finite-volume energies. However, a more complicated analysis is required where $s$- and $p$-waves are fit simultaneously to
Eq.~\ref{e:det}. First preliminary results from an analysis of this type~\cite{d200} are shown in Fig.~\label{f:d200} on
a single ensemble of $N_{\rm f} = 2+1$ dynamical fermions at $m_{\pi} = 200{\rm MeV}$. 
Evidently, the inclusion of $s$- and $p$-wave dominated irreps provides a number of additional finite-volume energy levels 
which better constrain the $p$-wave Breit-Wigner fit parameters. The $s$-wave is modeled using the leading term in the 
effective range expansion from Eq.~\ref{e:eff}. 
\begin{figure}
  \includegraphics[width=0.7\textwidth]{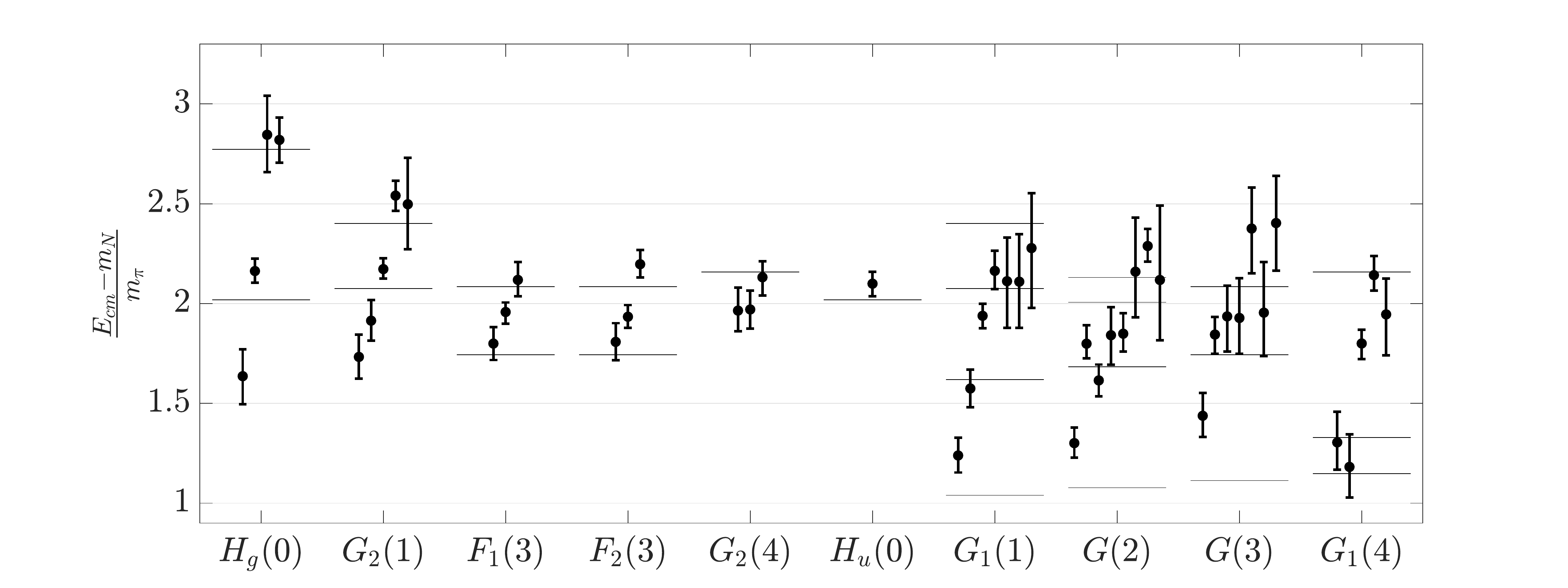}
  \includegraphics[width=0.29\textwidth]{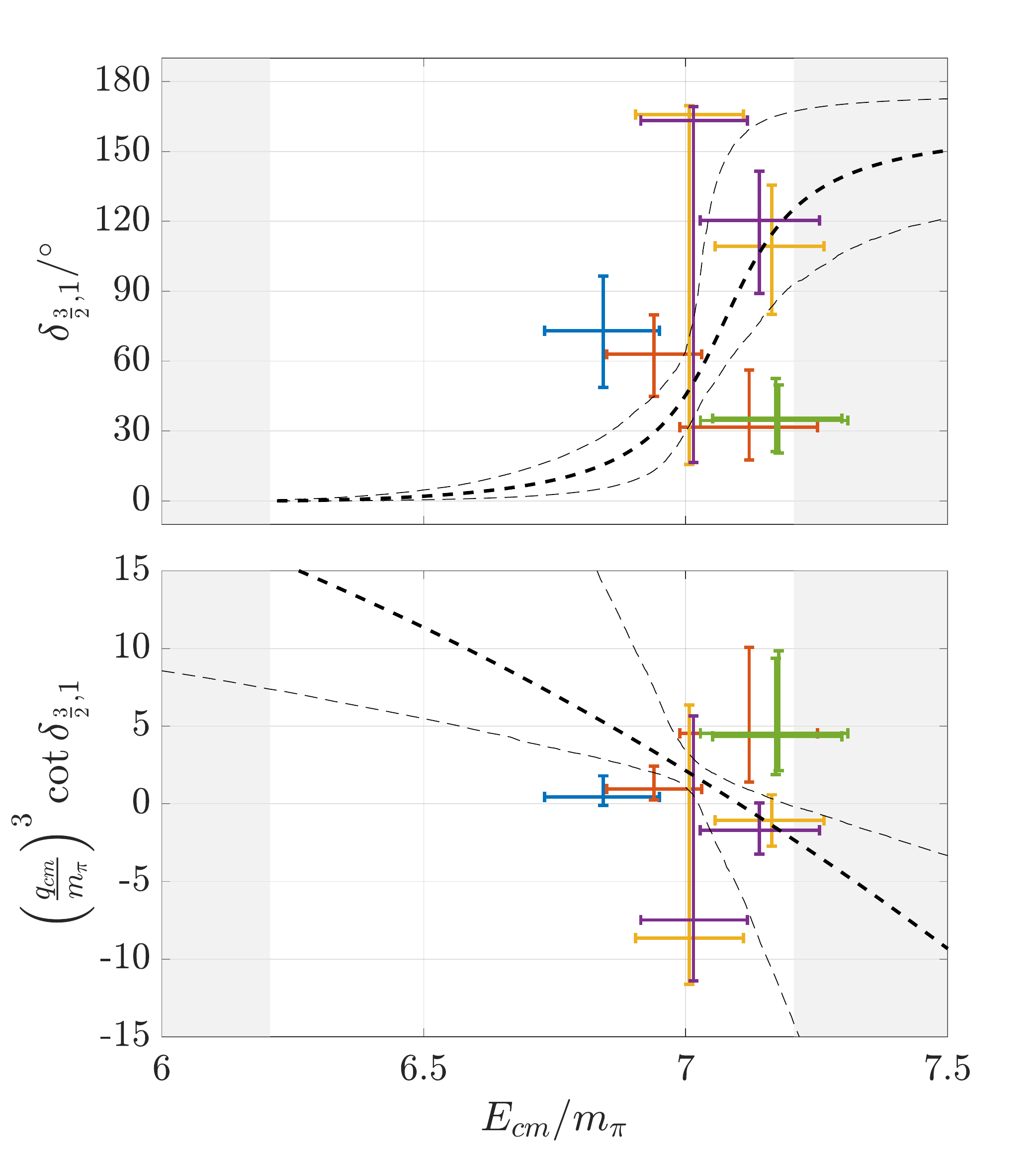}
  \caption{\label{f:d200} From Ref.~\cite{d200}, preliminary results for the elastic $I=3/2$ $p$-wave $N\pi$ 
  scattering amplitude on a single ensemble of $N_{\rm f}=2+1$ dynamical fermions at $m_{\pi}=200{\rm MeV}$. \emph{Left}:
  the finite volume spectra in irreps containing both $s$- and $p$-wave contributions. \emph{Right}: a plot of $p$-wave 
  dominated irreps for which the single-partial wave approximation is applied together with a Breit-Wigner fit to all levels
  where the $s$-wave is modeled by a constant.}
\end{figure} 

\section{Scattering from spectral functions}

Although it continues to be successful where applicable, the finite-volume formalism employed above has several shortcomings. 
Chiefly among them is the restriction of Eq.~\ref{e:det} to energies below three (or more) hadron thresholds, preventing the
study of many interesting systems including excited nucleon resonances. If the three-hadron formalism is fully developed 
and applied, finite-volume energy levels up to four (or more) hadron thresholds may be interpreted, but no general approach 
exists for levels above arbitrary inelastic thresholds. Further limitations of the finite-volume 
formalism include the inability to directly calculate inclusive rates such as the purely hadronic process 
$p+p \rightarrow  X$, where $X$ denotes a sum over all hadronic final states. 

An alternative approach to determining real-time scattering amplitudes from Euclidean correlation functions is motivated 
by expressing the two-point correlation function in Eq.~\ref{e:ct} as 
\begin{align}\label{e:spec}
C_{ij}(\boldsymbol{p},\tau) = \int_0^{\infty}dE \, \rho_{ij}(\boldsymbol{p},E) \, {\rm e}^{-E\tau}, 
\qquad \rho_{ij}(\boldsymbol{p},E) = \sum_n \delta(E-E_n)   \langle 0 | \hat{\cal O}_i(\boldsymbol{p}) | n \rangle \,
 \langle n | \hat{\cal O}^{\dagger}_j(0) | 0 \rangle
\end{align}
in terms of the spectral function $\rho_{ij}(\boldsymbol{p},E)$. These spectral functions are independent of the metric 
signature and therefore contain real-time information.  However, $C_{ij}(\boldsymbol{p},\tau)$
is determined from the lattice Monte Carlo calculation at a finite number of $\tau$, each with a statistical error. The 
reconstruction of $\rho_{ij}(\boldsymbol{p},E)$ from such data is therefore an ill-posed problem. 

A more modest goal is the determination of the smeared spectral function 
\begin{align}\label{e:smr}
 \rho^{\epsilon}_{ij}(\boldsymbol{p},E) =  \int_{0}^{\infty}d\omega \, \delta_{\epsilon}(E-\omega), \quad 
 \lim_{\epsilon \rightarrow 0^+}  \delta_{\epsilon}(x) = \delta(x)
\end{align}
where the smearing kernel $\delta_{\epsilon}(E-\omega)$ approaches a Dirac-$\delta$ function as the smearing 
width is reduced. While the smearing is a necessary limitation introduced by the nature of the problem, it is advantageous 
in bridging the gap between finite and infinite volume. Although unsmeared spectral functions are very different in 
finite and infinite volume, at finite (fixed) smearing width $\epsilon$ the infinite-volume limit is well defined. 
Furthermore, recent improvements~\cite{Hansen:2019idp} to the Backus-Gilbert algorithm~\cite{BG1,BG2} for spectral 
reconstruction enable the efficient determination of $\rho^{\epsilon}$ with an input functional form for the smearing kernel 
$\delta_{\epsilon}(E-\omega)$. These methods provide the exact smeared spectral function, but smeared with a (known) smearing kernel 
$\hat{\delta}_{\epsilon}(E,\omega)$ that is only approximately equal to the desired one $\delta_{\epsilon}(E-\omega)$. 
An ideal choice for the smearing kernel~\cite{Bulava:2019kbi, Poggio:1975af} is the real part of the standard 
$i\epsilon$-prescription for causal propagation
\begin{align}\label{e:ker}
\pi\delta_{\epsilon}(E-\omega) = {\rm Re}\, \frac{i}{E - \omega + i\epsilon} = \frac{\epsilon}{(E-\omega)^2 + \epsilon^2}.
\end{align}
which has desirable analyticity properties in $\epsilon$, unlike (for instance) a gaussian smearing kernel. 
Physical on-shell scattering amplitudes are then obtained from the ordered double limit 
$\lim_{\epsilon\rightarrow 0^+} \lim_{L\rightarrow \infty}$. This approach was proposed first for inclusive processes mediated 
by external currents in Ref.~\cite{Hansen:2017mnd} and extended to arbitrary scattering amplitudes in 
Ref.~\cite{Bulava:2019kbi}. 

\begin{figure}
  \includegraphics[width=0.49\textwidth]{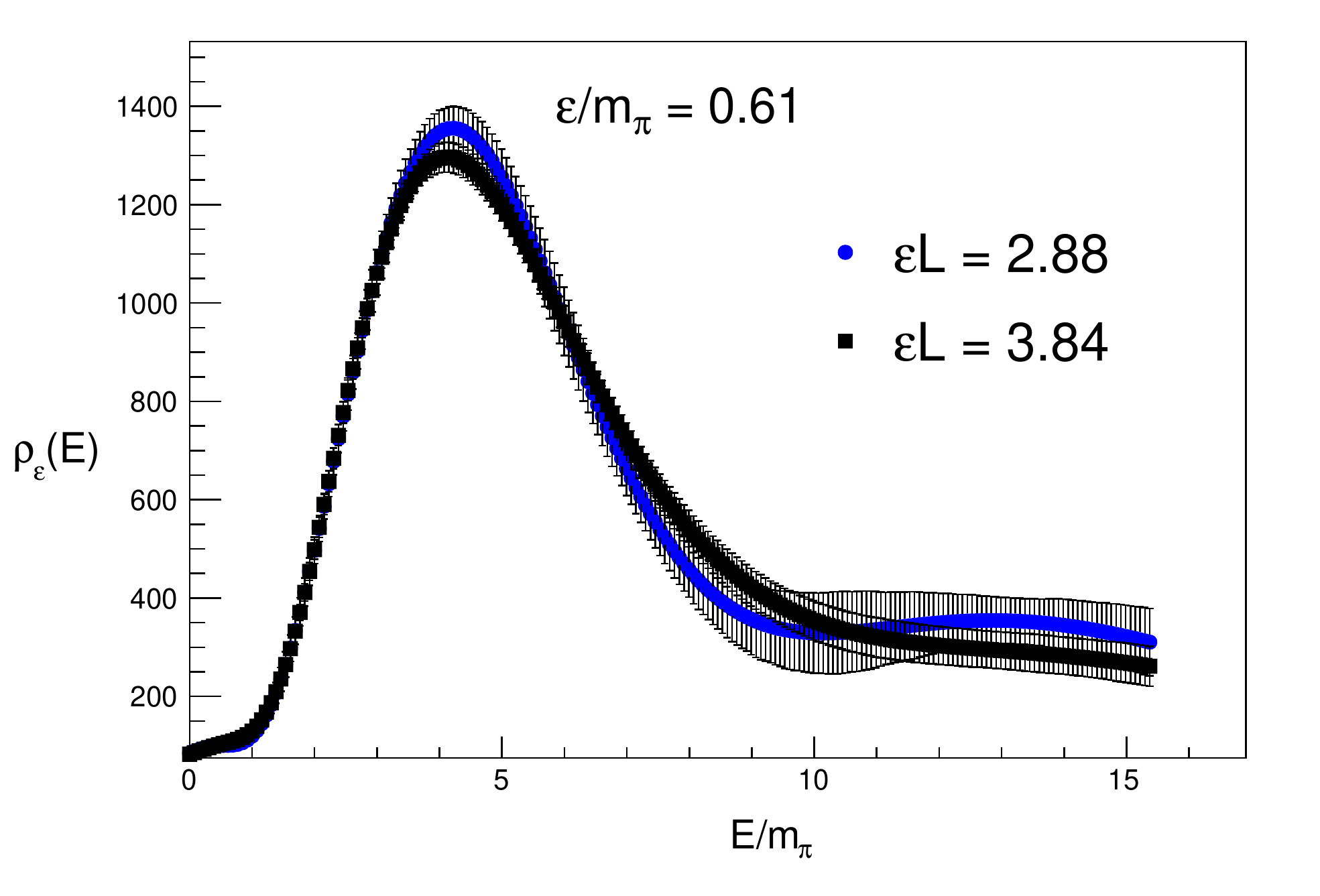}
  \includegraphics[width=0.49\textwidth]{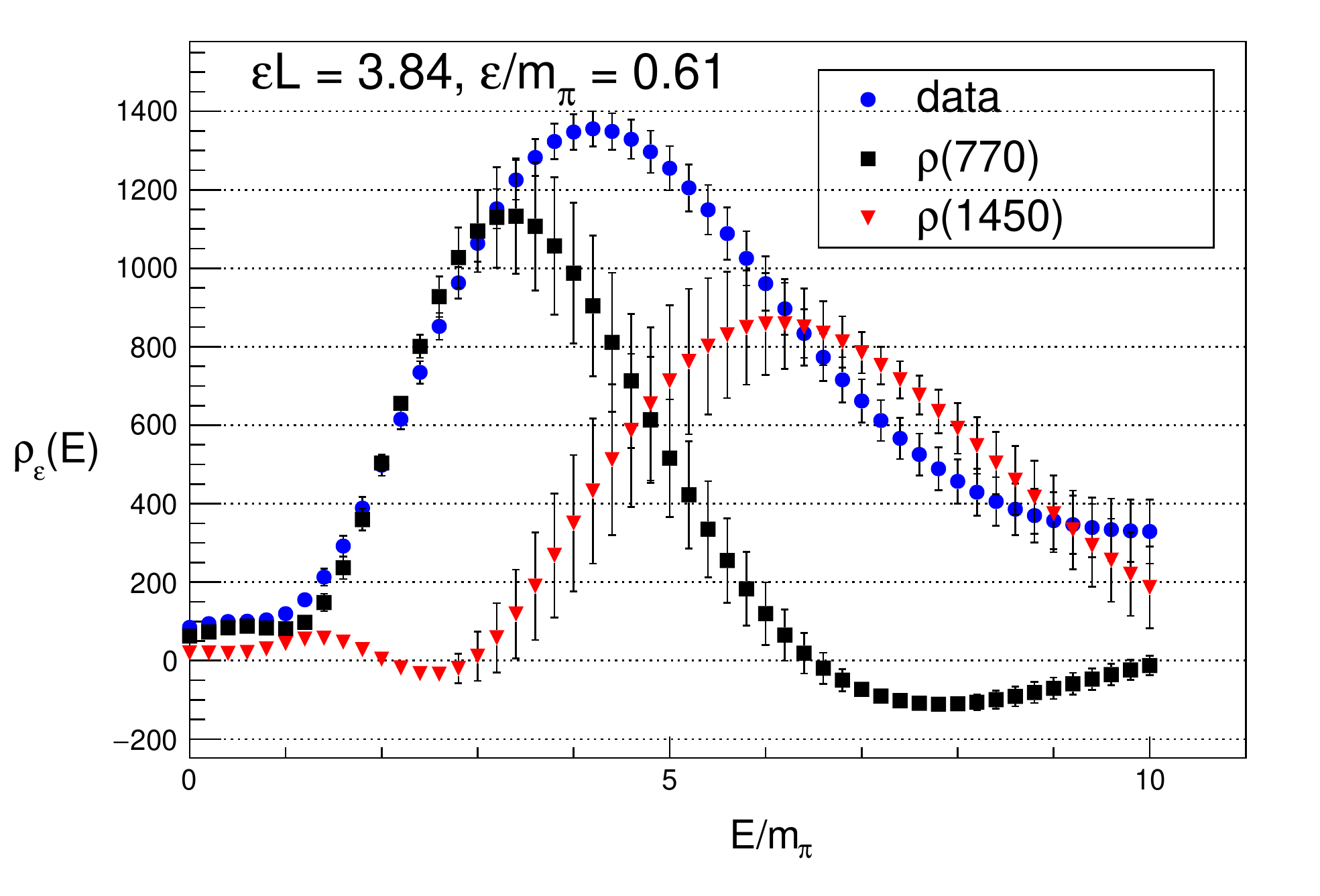}
  \caption{\label{f:incl} Spectral reconstruction of the smeared isovector vector 
  correlator from Ref.~\cite{Andersen:2018mau} for $N_{\rm f} = 2+1$ dynamical fermions at 
  $m_{\pi}=220{\rm MeV}$ and $a=0.086{\rm fm}$. \emph{Left}: $\rho_{\epsilon}(E)$ at two volumes illustrating the 
  rapid onset of the infinite volume limit at fixed $\epsilon$. \emph{Right}: Individual state contributions 
  of a fit to the data to Eq.~\ref{e:pole} with $N_{\rm pole}=2$ resulting in energies consistent with the $\rho(770)$ and the $\rho(1450)$. }
\end{figure}

As a illustrative application, consider the total rate for the inclusive process $\hat{J} \rightarrow {\rm hadrons}$, which $\hat{J}$ is 
an external current. An example of such a process is the ratio 
$R_{\rm had}(s) = \sigma({\rm e}^+{\rm e}^- \rightarrow {\rm hadrons})/\frac{4\pi\alpha_{\rm em}(s)}{3s}$ 
with the electromagnetic external current, which is relevant for the phenomenology of 
$(g-2)_{\mu}$~\cite{Jegerlehner:2009ry}. Fictitious inclusive processes can also be considered, in which 
the currents do not correspond to real-world external probes, but can be an arbitrary hadronic interpolator 
$\hat{\cal O}_{\rm had}$. In order for such processes to act as 
novel probes of QCD, these hadronic interpolators must be renormalized to possess a well-defined continuum limit. 
The renormalization of arbitrary hadron interpolators can be accomplished either using the Wilson 
flow~\cite{Luscher:2010iy,Luscher:2013cpa} or 
the approach of Ref.~\cite{Papinutto:2018ajw}, but indicative preliminary results are shown with a smeared 
unrenormalized $I=1$ vector current in Fig.~\ref{f:incl}. There the inclusive 
rate $\sigma({\cal O}_{\rm had}\rightarrow {\rm hadrons})(E) \propto \lim_{\epsilon \rightarrow 0^+} \rho_{\epsilon}(E)$ is 
determined at finite $\epsilon$ and multiple volumes using the 
freely-available lattice correlator data from Ref.~\cite{Andersen:2018mau}. 

Two different spatial volumes 
illustrate the rapid approach to the $L\rightarrow \infty$ limit at finite $\epsilon$, so that an infinite-volume 
\emph{ansatz} for $\rho_{\epsilon}(E)$ is appropriate. In this channel peaks associated with the $\rho(770)$ resonance as well 
as excited $I^G(J^{P}) = 1^+(1^-)$ resonances are expected. However, the smearing width $\epsilon = 0.61m_{\pi} =
135{\rm MeV}$ is larger than the typical width of these states so that an \emph{ansatz} that treats them as infinitesimally narrow
is appropriate. To fit $\rho_{\epsilon}(E)$, we employ the model 
\begin{align}\label{e:pole}
\rho_{\epsilon}(E) = \sum_{n=1}^{N_{\rm pole}} A_n \, \delta(E-E_n)  
\end{align}
to describe the data in Fig.~\ref{f:incl}. The $N_{\rm pole}=1$ form does not provide a good description of the data, 
resulting in an (uncorrelated) $\chi^2/{\rm d.o.f.} = 42.8$, 
but using the four-parameter $N_{\rm pole} = 2$ \emph{ansatz} reduces this to $\chi^2/{\rm d.o.f.} = 0.65$. The 
resultant energies in this fit are $E_1=740(30){\rm MeV}$ and $E_2 = 1394(109){\rm MeV}$ which are consistent with 
the experimentally determined masses of the $\rho(770)$ and $\rho(1450)$.  

While this inclusive example models the smeared total rate at finite $\epsilon$, exclusive amplitudes are can be 
recovered in the $\epsilon\rightarrow 0^+$ limit using the LSZ reduction approach of Ref.~\cite{Bulava:2019kbi}. As a first 
demonstration
of using spectral functions to obtain exclusive amplitudes~\cite{O3}, consider the $1+1$-dimensional $O(3)$ model employed in 
Ref.~\cite{Luscher:1990ck}. Like QCD, this model is asymptotically 
free and has a mass gap, but the elastic two-to-two scattering amplitude is known analytically~\cite{Zamolodchikov:1977nu} 
so it is an ideal test case. Furthermore the global $O(3)$ symmetry is reminiscent of isospin symmetry in QCD, so that 
two-to-two scattering proceeds in one of three `isospin' channels with $I=0,1,2$.  

In order to calculate the two-to-two scattering amplitude using the LSZ reduction approach, the connected 
Euclidean four-point temporal correlation function is computed and the two outer time separations are taken 
asymptotically large. The resultant correlation function has a single Euclidean time argument, to which the 
spectral reconstruction of Eqs.~\ref{e:spec} and~\ref{e:smr} is applied to obtain 
the smeared spectral function $\rho_{\epsilon(E)}$. The 
two-to-two scattering amplitude $M_2(E^*)$ is then computed from the ordered double limit
\begin{align}\label{e:222}
M_2(E^*) = \lim_{\epsilon \rightarrow 0^+} \lim_{L \rightarrow \infty} M^{\epsilon}_2(E^*), \qquad 
M^{\epsilon}_2(E^*) = Z^{-1}\, \epsilon^2\, \rho_{\epsilon}(E^*)  
\end{align}
where $E^*$ is the total (lab frame) energy and $Z$ the interpolator overlap factor. Note the factor of $\epsilon^2$ which 
performs the `amputation' of the on-shell pole. Since only the real part of the amplitude is considered here, the 
smearing kernel from Eq.~\ref{e:ker} is employed.   

\begin{figure} 
  \includegraphics[width=0.49\textwidth]{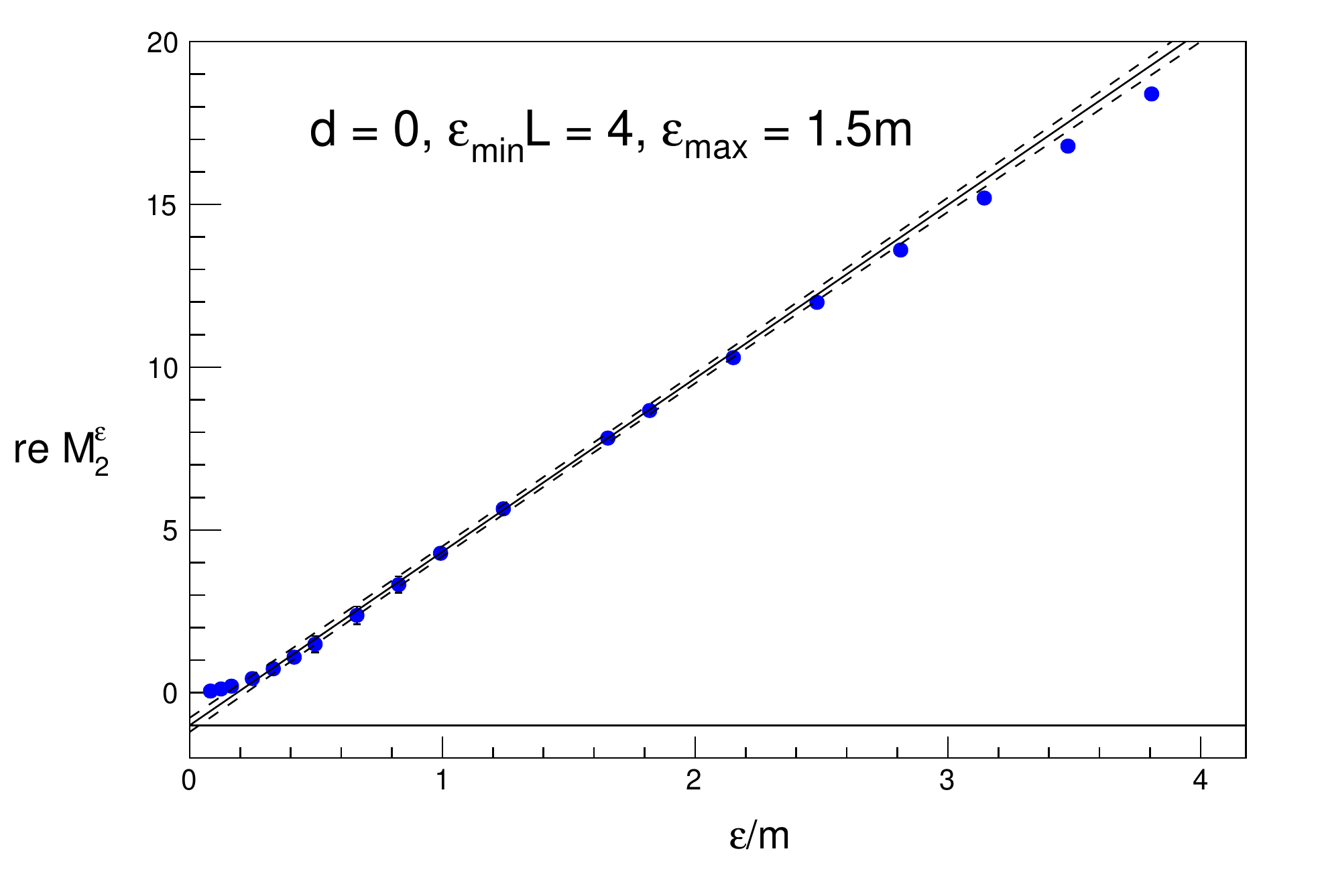}
  \includegraphics[width=0.49\textwidth]{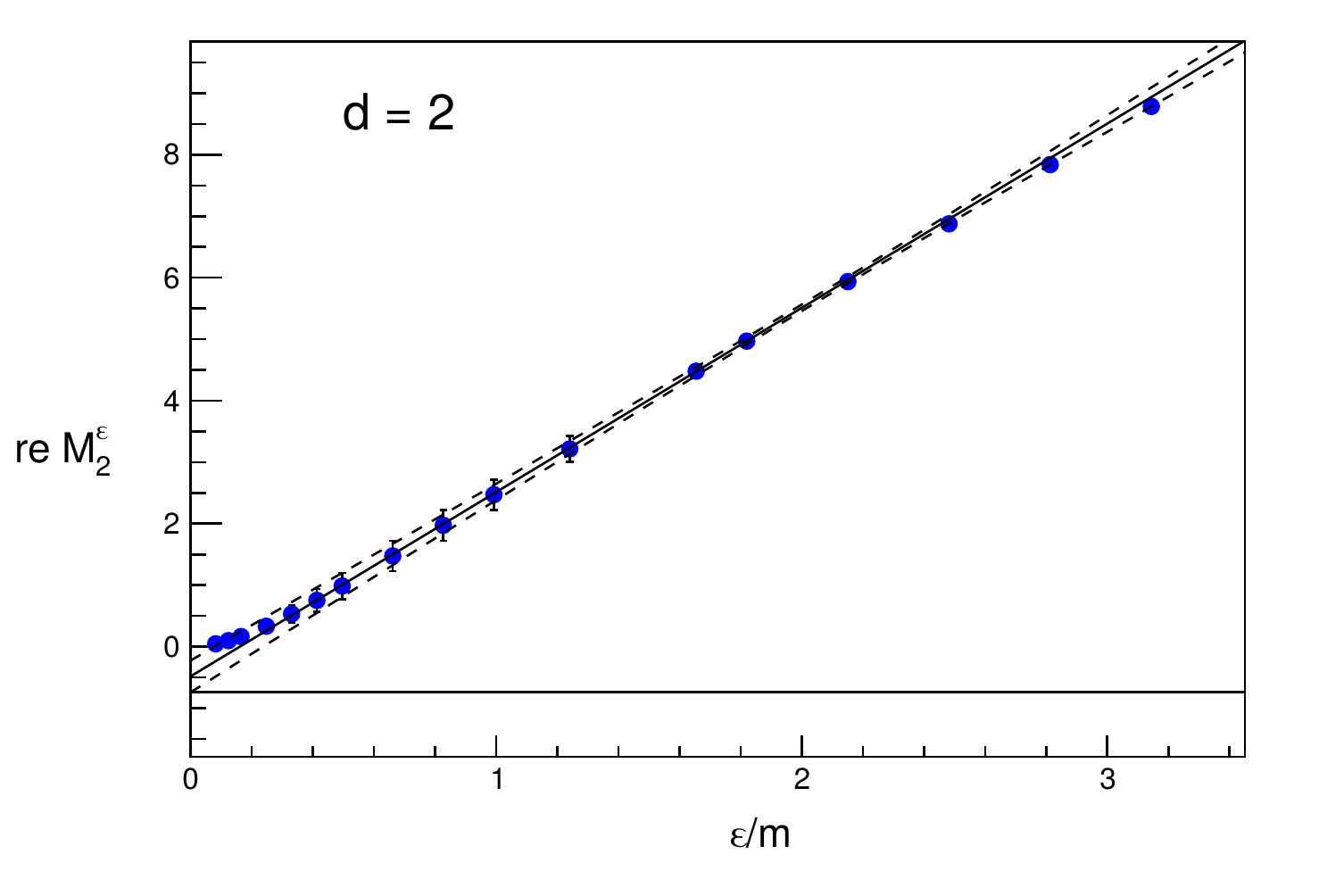}
  \caption{\label{f:lsz} Preliminary results for the real part of the elastic $I=2$ two-to-two scattering amplitude in the 1+1 dimensional
  $O(3)$ model with $mL = 19.4$. For zero total momentum, results for $M_2^{\epsilon}(E^*)$ from Eq.~\ref{e:222} 
  are shown for zero units of relative momentum (\emph{left}) and two units of
  relative momentum (\emph{right}). For each case, the order double limit is performed by extrapolating linearly to 
  $\epsilon \rightarrow 0$ over the range $[\epsilon_{\rm min}, \epsilon_{\rm max}]$. The horizontal 
  solid line corresponds to the exact results from Refs.~\cite{Luscher:1990ck,Zamolodchikov:1977nu}, which is consistent 
  with the extrapolated values. Deviations from linearity are evident for $\epsilon < \epsilon_{\rm min}$ illustrating 
  the importance of the ordered double limit. }
\end{figure}
The double limit in Eq.~\label{e:222} is performed numerically in Fig.~\ref{f:lsz} at a fixed $mL = 19.4$ by extrapolating 
linearly to $\epsilon=0$ using data in the range $\epsilon_{\rm min} < \epsilon < \epsilon_{\rm max}$, with 
$\epsilon_{\rm min} L = 4$ to remain safe from finite-volume effects and $\epsilon_{\rm max} = 1.5m$ to remain in the 
linear regime. As is evident in Fig.~\ref{f:lsz}, this extrapolation results in scattering amplitudes consistent with 
the known analytic result. Deviations from linearity are also apparent in Fig.~\ref{f:lsz} for 
$\epsilon < \epsilon_{\rm min}$, which illustrate the need for the appropriate ordered double limit.  

\section{Conclusions}

This review surveys the current state of lattice QCD calculations of meson-nucleon scattering amplitudes, which is 
an emerging subfield.  
The finite-volume formalism of Eq.~\ref{e:det} has been successful in first calculations of two-to-two amplitudes below 
three or more hadron thresholds. 
These include near-threshold studies which determine parameters of the effective range expansion as well as first 
studies of the $\Delta(1232)$ resonance across the entire elastic region. 
All meson-nucleon calculations to date have also treated only elastic scattering, although
Eq.~\ref{e:det} applies also to coupled two-hadron scattering channels which are relevant for the 
$\Lambda(1405)$ resonance. Furthermore, extension of the finite-volume formalism to treat meson-meson-baryon
 channels and the corresponding lattice QCD computation of the spectra 
 have not yet been completed, although first calculations of three-pion scattering amplitudes are 
encouraging~\cite{Horz:2019rrn,Blanton:2019vdk,Mai:2019fba}. 
Nonetheless, the range of impact of the two-to-two scattering amplitudes can be increased by using lattice QCD scattering 
data as an input to effective field theories and EFT-inspired models. Work in this direction with nucleons has already 
been performed~\cite{Lutz:2018cqo,Liu:2016uzk}.

In general progress in meson-baryon scattering on the lattice has been slower than meson-meson 
calculations due to several difficulties, such as the increased signal-to-noise problem, complications in correlator 
construction due to the extra quark, and complications in the finite-volume formalism from the inclusion of non-identical
particles with non-zero overall spin. 
Despite these difficulties, it is reasonable to assume that
 (in analogy with the meson-meson sector) precise results for elastic amplitudes and coupled meson-baryon scattering 
 channels will be produced in the near future. Making further comparisons with meson scattering amplitudes, 
 it is likely that calculations of meson-baryon amplitudes mediated by external electroweak 
 currents $\hat{J}_{\rm ew}$ are also possible, such as $N  + J_{\rm ew} \rightarrow  N+\pi$ and 
 $N+\pi+J_{\rm ew} \rightarrow N+\pi$, which have a number of phenomenological applications. There has been substantial 
 progress on analogous amplitudes in the meson-meson sector, such as $\gamma^* \rightarrow \pi\pi$ in 
 Refs.~\cite{Feng:2014gba,Andersen:2018mau} and $\pi + \gamma^* \rightarrow \pi\pi$ in 
 Refs.~\cite{Briceno:2016kkp,Alexandrou:2018jbt}. 

The final topic discussed in this review is a novel approach to computing scattering amplitudes in lattice QCD simulations 
based on spectral functions. This does not employ the finite volume whatsoever and as such requires 
large volumes to saturate the $L\rightarrow \infty$ limit. Because of this, larger volumes 
are required than in typical lattice QCD simulations, although some preliminary indicative results for inclusive 
decays on existing lattice QCD data are encouraging. The LSZ reduction approach of Ref.~\cite{Bulava:2019kbi} 
has not yet been employed in lattice QCD but preliminary results in a 1+1 dimensional toy model reproduce the known 
elastic two-to-two scattering amplitude. Overall, if feasible the infinite-volume formalism based on spectral functions 
has the potential to circumvent many of the limitations present when using the finite-volume, such as partial wave and 
coupled-channel mixing as well as the difficulty in going above arbitrary inelastic thresholds. These issues are of course 
relevant when studying excited nucleon resonances. However, the degree to which 
Refs.~\cite{Bulava:2019kbi} and~\cite{Hansen:2017mnd} can be applied to lattice QCD simulation data is an open question.    


\nocite{*}
\bibliography{latticen}

\end{document}